%% file: Main_document.tex
\RequirePackage{rotating}

\documentclass{jfm}
\usepackage{graphicx}
\usepackage{epstopdf, epsfig}
\usepackage{amsmath}
\usepackage{xargs}                      % Use more than one optional parameter in a new commands
\usepackage[pdftex,dvipsnames]{xcolor}  % Coloured text etc.
\usepackage{color}
\usepackage{xcolor}
\usepackage{tikz}
\usepackage{import}
\usepackage{amsfonts}
\usepackage{amssymb}
\usepackage{graphicx}
\usepackage{pgfplots}
\pgfplotsset{compat=1.6}
\usepgfplotslibrary{fillbetween}
\usetikzlibrary{patterns}
\usetikzlibrary{plotmarks}
\usetikzlibrary{matrix}

\usepackage{layouts}
\usepackage[colorlinks = true,allcolors = {blue}]{hyperref}
\usepackage{multirow}
\usepackage{subcaption}
\usepackage{siunitx}
\usepackage{tabularx}

\usepackage{parnotes}
\usepackage{makecell}
\usepackage{rotating}

\newcommand{\perc}[1]{ \SI{#1}{\percent} }
\newcommand{\e}[1]{\times 10^{#1}}
\definecolor{darkgreen}{rgb}{0.13, 0.55, 0.13}

\newcommand{\add}[1]{\textcolor{black}{#1}}

\shorttitle{Bubbly DR using a hydrophobic IC in TC turbulence}
\shortauthor{P. A. Bullee and others} %, Verschoof, Bakhuis, Ezeta, Huisman, Sun, Lammertink, Lohse }

\title{Bubbly drag reduction using a hydrophobic inner cylinder in Taylor--Couette turbulence}

\author{Pim A. Bullee\aff{1,2},
  Ruben A. Verschoof\aff{1},
  Dennis Bakhuis\aff{1},
  Sander~G.~Huisman\aff{1},
  Chao~Sun\aff{4,1}
	  \corresp{\email{chaosun@tsinghua.edu.cn}},
  Rob~G.~H.~Lammertink\aff{2}
	  \corresp{\email{r.g.h.lammertink@utwente.nl}},
\and Detlef~Lohse\aff{1,3}
 \corresp{\email{d.lohse@utwente.nl}}
 }

\affiliation{
\aff{1}Physics of Fluids, Max Planck Center Twente for Complex Fluid Dynamics, MESA+ Research Institute and J. M. Burgers Centre for Fluid Dynamics, University of Twente, P.O. Box 217, 7500 AE Enschede, The Netherlands
\aff{2}Soft matter, Fluidics and Interfaces, MESA+ Research Institute, University of Twente, P.O. Box 217, 7500 AE Enschede, The Netherlands
\aff{3}Max Planck Institute for Dynamics and Self-Organization, Am Fassberg 17, 37077 G\"{o}ttingen, Germany
\aff{4}Center for Combustion Energy and Department of Energy and Power Engineering, Tsinghua University, 100084 Beijing, China
}

\begin{document}

\maketitle

\begin{abstract}
In this study we experimentally investigate bubbly drag reduction in a highly turbulent flow of water with dispersed air at $5.0 \e{5} \leq \Rey \leq 1.7\e{6}$ over a non-wetting surface containing micro-scale roughness. To do so, the Taylor--Couette geometry is used, allowing for both accurate global drag and local flow measurements. The inner cylinder -- coated with a rough, hydrophobic material -- is rotating, whereas the smooth outer cylinder is kept stationary. The crucial control parameter is the air volume fraction $\alpha$ present in the working fluid. For small volume fractions ($\alpha < \SI{4}{\percent}$), we observe that the surface roughness from the coating increases the drag. For large volume fractions of air ($\alpha \geq \SI{4}{\percent}$), the drag decreases compared to the case with both the inner and outer cylinders uncoated, i.e. smooth and hydrophilic, using the same volume fraction of air. This suggests that two competing mechanisms are at place: on the one hand the roughness invokes an extension of the log-layer -- resulting in an increase in drag -- and on the other hand there is a drag-reducing mechanism of the {hydrophobic} surface interacting with the bubbly liquid. The balance between these two effects determines whether there is overall drag reduction or drag enhancement. For further increased bubble concentration $\alpha = \SI{6}{\percent}$ we find a saturation of the drag reduction effect. Our study gives guidelines for industrial applications of bubbly drag reduction in hydrophobic wall-bounded turbulent flows.

%% Conference abstract
%We investigate the drag of a highly turbulent flow over a non-wetting surface of micro-scale roughness. The Taylor--Couette geometry is used, allowing accurate drag and flow measurements. In this study, the inner cylinder is rotating, whereas the outer cylinder is kept stationary. The gap-width based Reynolds number is $\mathcal{O}(10^6)$. The inner cylinder is coated with a rough, hydrophobic material. whereas the outer cylinder is kept smooth. We vary the void fraction of air $\alpha$ present in the working fluid to introduce bubbles to the flow. For smaller volume fractions of air, up to $\alpha \leq 2\%$, we observe that the increased surface roughness from the coating increases the drag. For larger fractions of air, $\alpha > 2\%$, the drag decreases compared to a smooth hydrophilic, uncoated cylinder using the same volume fraction of air. This suggests that two mechanisms play a role: the roughness invokes a shift in the log-layer --- resulting in an increase in drag ---  and the more effective drag-reducing mechanism of the {hydrophobic} surface. The balance between these two effects determines whether bubble drag reduction is more effective when using a {hydrophobic} surface compared to using a smooth hydrophilic surface.

\end{abstract}

\begin{keywords}
% picked from: https://www.cambridge.org/core/services/aop-file-manager/file/577245f575fc11ff73cc3197/jfm-keywords.pdf
Taylor--Couette flow, drag reduction, coating
\end{keywords}
%
\include{introduction}
\include{experimental}
\include{results}
\include{summary_conclusions}

\section*{Conflict of interest}
The authors declare no conflicts of interests.

\begin{acknowledgments}
\section*{Acknowledgements}
We would like to thank Gert-Wim Bruggert, Martin Bos and Bas Benschop for their continuous technical support over the years with the T$^3$C. We thank Ineke Punt and Jan van Nieuwkasteele for their help with the {hydrophobic} coating and all related issues. We acknowledge stimulation discussions with Pieter Berghout on roughness, and on numerous of other related topics with Jeffery Wood. We thank Arne te Nijenhuis for his help in the lab and Rodrigo Ezeta for his help with the PIV experiments. The wetting experiments with TritonX where done together with Remco van der Maat as part of his Bachelor graduation assignment. This research is supported by the project ``GasDrive: Minimizing emissions and energy losses at sea with LNG combined prime movers, underwater exhausts and nano hull material'' (project 14504) of the Netherlands Organisation for Scientific Research (NWO), domain Applied and Engineering Sciences (TTW). R.A.V. acknowledges NWO-TTW (project 13265). S.H. acknowledges financial support from MCEC. C.S. and D.B. acknowledge financial support from VIDI grant No. 13477, and the Natural Science Foundation of China under grant no. 11672156.
\end{acknowledgments}

\bibliographystyle{jfm}
\bibliography{biblio}

\end{document}

%% file: introduction.tex
% !TEX root = Main document.tex
\section{Introduction}
Skin friction drag reduction (DR) in turbulent flow is a topic of research that is relevant for many industrial applications. In particular, the maritime industry may benefit from this, since reducing fuel consumption by only a few percent will lead to significant cost savings and reduction of pollutant emission~\citep{vandenBerg2007,Ceccio2010,Murai2014,Park2014,Gose2018}.

In this work we combine {hydrophobic} surfaces with two phase flow to study drag reduction, \add{a combination that, to our best knowledge, has not often been studied before, especially not at the high Reynolds numbers $\Rey$ of up to $1.8\e{6}$ that we reach. The physics behind this combination is interesting, since both hydrophobic surfaces and (air) bubble injection have shown individually to decrease the skin friction drag. At the same time, by increasing the amount of gas in the liquid, the effectivity and life span of a drag reducing superhydrophobic surface can be increased~\citep{Lv2014, Xiang2017}. Compared to a hydrophilic surface, gas bubbles that impact a hydrophobic surface are more likely to attach to the surface and form a lubricating layer~\citep{Kim2017}. Although the wall shear stress in our setup is much larger than what the bubbles in the work of~\citet{Kim2017} are exposed to, a possible result is that the number of bubbles close to the wall increases, which is beneficial for bubbly DR. A set of experiments of two-phase flow over a hydrophobic plate up to $\Rey = 5000$ by~\citet{Kitagawa2019} showed two groups of bubbles. One group of medium-sized free bubbles, and a group of small wall-adhered bubbles, that coalesce into large bubbles. Since the bubbles that stick to the plate change the flow close to the plate, they suggest that the hydrophobic plate is likely to experience more friction drag. Based on this reasoning, they suggest that these results should be carefully considered, when air bubble behaviour is controlled using functionalized (hydrophobic) surfaces in bubbly DR applications~\citet{Kitagawa2019}. Hence, the two methods of drag reduction (bubbly and with hydrophobic surfaces) will influence one-another. However, it is yet unknown whether this is positive or negative for the total combined drag reduction and we want to find this out in this paper.} \\

We explore the difference in skin friction coefficient between two types of surfaces: a very smooth hydrophilic surface and a more rough {hydrophobic} surface. The {hydrophobic} surface is a sheet of porous polypropylene material, commercially available in large quantities. Representative to more practical applications, it has a sponge-like isotropic geometry of distributed (roughness) length scales formed by the porous structure. To study the fully developed turbulence typical for maritime applications, it is desirable to experimentally achieve high Reynolds numbers, and have both the bulk flow and boundary layer in a state of turbulence. To this end, we use the Twente Turbulent Taylor--Couette facility (T$^3$C) described in~\citet{VanGils2011}, of which the inner cylinder is made {hydrophobic} using the porous polypropylene material. This closed system, with an exact energy balance between input (driving of the flow) and output (viscous energy dissipation), allows for accurate measurement of global drag. Due to its excellent optical accessibility, this can be combined with local flow measurements, for instance using particle image velocimetry (PIV), as well as visualisations of the flow structure and the {hydrophobic} surface using (high-speed) imaging techniques. Air bubbles are introduced to the working liquid to demonstrate the drag reducing effect of the {hydrophobic} inner cylinder. This combination of the T$^3$C with a SH inner cylinder and air bubbles in the working fluid, enables us to study {hydrophobic} bubbly drag reduction at industrially relevant high Reynolds numbers in a well controlled condition, giving a better understanding of the mechanisms involved.

The paper is organized as follows: In chapter 2 we give an extensive overview of prior work on bubbly drag reduction, drag on {hydrophobic} surfaces, and drag enhancement of rough walls, as all these effects are crucial to understand the competing effects explored in this paper. In chapter 3 the experimental methods are described. Chapter 4 presents the results and discusses them. The paper ends with conclusions.

\section{Overview over prior work on bubbly drag reduction and (super)hydrohobic surfaces}

\subsection{Drag reduction with {hydrophobic} surfaces}
Superhydrophobic surfaces are typically created by combining a hydrophobic chemistry (resulting in low surface energy) with micro or nanoscale asperities on the surface~\citep{Li2007}. The top of these asperities are in contact with the liquid, while air is captured between the asperities. This effectively reduces the solid-liquid contact area, partially replacing it with a gas-liquid interface, that locally changes the no-slip boundary condition to a shear-free boundary condition.
The gas-liquid interface is supported by the capillary forces, which in general are larger for hydrophobic materials compared to hydrophilic materials of equal geometry. Dependent of chemistry and geometry, a gas-liquid interface can collapse under influence of a pressure or shear force, and transitions into a thermodynamically favoured wetting state. Various types of asperities exist, ranging from structures such as pillars and ridges to pyramids and mushroom-like shapes~\citep{Peters2009,Qi2009,Park2014,Domingues2017}. However, such well-defined shapes are expensive and time-consuming to produce. Therefore, larger areas of SH surfaces~($>\SI{100}{\square\centi\metre}$) usually have a random roughness structure~\citep{Hokmabad2016}. We refer the reader to the review article by~\cite{Li2007} for a broader introduction to SH surfaces.\\

An overview of various experimental and numerical studies in the laminar and low Reynolds number ($\Rey$) turbulent regime is given in the review article by~\cite{Rothstein2010}. Under laminar flow conditions, the behaviour of SH surfaces is typically studied in microchannels. Drag reduction is then quantified by defining a slip length, a slip velocity or by a decrease in pressure drop over the channel~\mbox{\citep{Tsai2009,Haase2013,Park2015}}. As many industrial flows are highly turbulent, it is crucial to study the behaviour of such surfaces in the high Reynolds number flow regime. For marine vessels for example, Reynolds numbers are of the order of~$\Rey = \mathcal{O}(10^9)$.

Superhydrophobic DR in laminar flow only depends on the geometry of the asperities on the surface that set the slip length and determine the slip velocity. For turbulent flows, SH drag reduction also depends on the Reynolds number~\citep{Park2013}. With increasing $\Rey$, the thickness of the viscous sublayer decreases, which is the most relevant length scale when comparing the geometric features of the superhydrophobic surface~\citep{Daniello2009}. In the near-wall region inside the boundary layer of a turbulent flow, the momentum transfer is dominated by molecular interactions, whereas the role of turbulent momentum transfer is negligible. In other words, viscous stress dominates over Reynolds stress. Altering this region affects the entire boundary layer and hence the drag. The outer edge of the viscous sublayer is typically given by a distance $y_\text{vsl} = 5 \nu / u_\tau = 5 \delta_\nu$ from the wall, where $\nu$ is the kinematic viscosity, and $u_\tau = \sqrt{\tau_w / \rho}$ the friction velocity for wall shear stress $\tau_w$ and density $\rho$. The viscous length scale $\delta_\nu = \frac{\nu}{u_\tau}$ is the usual scaling parameter for nondimensionalization to viscous wall units, indicated by a superscript `+', e.g. $y^+ = y/\delta_\nu$.

In laminar flow, DR is a direct result of the shear-free (slip) boundary condition. An additional effect matters in turbulence, where near-wall turbulent structures are suppressed due to the slip boundary condition, resulting in additional DR~\citep{Park2013}. The numerical work of \cite{Park2013} showed that DR increases with the slip length $b^+$, defined as the length below the surface where the extrapolated velocity profile reaches zero. When $b^+ \gtrapprox 30-40$, the drag is not further affected by an increase of $b^+$. This length scale corresponds to the outer edge of the buffer layer $5 < y^+ < 30$~\citep{Pope}, where streamwise near-wall vortical structures primarily reside~\citep{Park2013}. Both observations point in the direction that these near wall structures are very important for the larger DR that is found for turbulent flows over SH surfaces compared to laminar flow over SH surfaces~\citep{Park2013}. \add{In the work of~\citet{Rastegari2018} similar conclusions were drawn. A balance was found between the drag reducing mechanisms of superhydrophobic microgrooves and riblets in the form of a slip velocity together with weakened Reynolds shear stress and near wall vortical structures on the one hand, and a drag increase from the interactions between the microtextures and the flow on the other hand.} Results from experiments by \cite{Daniello2009} suggest a critical Reynolds number that prompts the onset of DR, which corresponds to the transition to turbulent flow. For their system of streamwise-aligned SH ridges in channel flow, they find no DR in the laminar regime, whereas after the flow has transitioned to turbulent flow, significant drag reduction was found. Hence, the physics behind the onset of DR must be related to the structure of the wall-bounded turbulent flow~\citep{Daniello2009}.

We divide the literature on turbulent flow over {hydrophobic} surfaces in two regimes: low (but still turbulent) $\Rey$ turbulence ($\Rey < 10^5$) and high $\Rey$ turbulence ($\Rey > 10^5$). In these regimes, a difference between single-phase and two-phase flow can be made, although most of the research so far has focussed on single-phase flow. Note that in single-phase flow, i.e., when no air is actively added to the working liquid, air might be trapped by the SH surface when the surface is submerged in the working liquid. In two-phase flow, gas is actively dispersed by (for instance air) bubble injection to the working liquid.

Different design rules are suggested in literature for optimal size and spacing of the geometrical features forming the SH surface. In the low $\Rey$ turbulence regime, authors mainly seem to use, or suggest to use, surfaces with pillar/ridge spacing $w^+ > 1$, or with a roughness parameter $k^+>1$. For the high $\Rey$ turbulence regime, however, the opposite is the case: suggested is $w^+ < 1$, or $k^+ < 1$. \add{The study by~\citet{Gose2018} suggests to not only use the normalized roughness $k^+$ to predict the drag reducing properties of a superhydrophobic surface, but to also include the contact angle hysteresis measured at a pressure higher than atmospheric pressure. This is done to simulate the large pressure fluctuations and high shear rates generated by high Reynolds number flows~\citep{Gose2018}. The roughness of the superhydrophic surfaces they studied varied between $k^+ = 0.2$ and $k^+ = 4.5$, with corresponding drag reduction changing from \perc{-90} to \SI{90}{\percent}. Specifically around $\text{DR} = \perc{0}$, the trend of increasing DR with decreasing $k^+$ is absent, showing drag reduction for one surface with $k^+ = 1$ and an increase of drag for another surface with $k^+ < 1$. When $k^+$ was scaled with the roughness parameter and the wetted area fraction, or the high-pressure contact angle hysteresis (\SI{370}{Pa} for a \SI{250}{\nano \litre}), the DR data collapsed to a single curve~\citep{Gose2018}. }

An overview of the different surface parameters found in literature focussing on DR with SH surfaces is shown in table~\ref{table:DRtable}. \\
\begin{sidewaystable}
\begin{tabularx}{\textwidth}{lllllll}
\hline
Author & Phase & Flow type & SH surface & \Rey & DR$_\text{max}$ & Surface roughness\\
\cite{Srinivasan2011} & single & Taylor--Couette & random & $1.6\times10^3-8.0\times10^4$ & \SI{22}{\%} & \makecell[tl]{$k^+ = 1.4$\parnote[a]{derived from data in paper}\\ $w^+ = 2.5$\parnotemark{a} } \\
\cite{Rastegari2018} & single & channel (LB) & longitudinal grooves & $3.6\e{6}$ & \SI{61.1}{\%} & $w^+=8$\parnote[b]{normalized with value for no-slip surface} \\
\cite{Daniello2009} & single & channel & ridges & ${3.0}\times10^3-{6.0}\times10^3$ & \SI{50}{\%} & $w^+ > 1$\parnote[c]{optimal value obtained from parameter sweep} \\
\cite{Martell2009}   & single & channel (DNS) & \makecell[tl]{ridges \\ posts} & $4.2\e{3}$\parnotemark{a} & \SI{40}{\%} & \makecell[tl]{$w^+$ as large \\ as possible} \\
\cite{Rosenberg2016} & single & Taylor--Couette & triangular ridges & $6.0\e{3}-9.0\e{3}$ & \SI{10}{\%} & \makecell[tl]{$k^+ = 3.9$ \\ $w^+ = 5.6$ } \\
\cite{vanBuren2017} & single & Taylor--Couette & ridges & $6.0\e{3}-1.0\e{4}$ & \SI{9}{\%} & $w^+ = 35$\\
\cite{Park2013} & single & channel (DNS) & ridges & $4.2\e{3} - 1.7\e{4}$ & \SI{90}{\%} & $w^+ = 100$\parnotemark{c} \\
\cite{Gose2018} & single & channel & random & $1.0\e{4}-3.0\e{4}$ & \SI{90}{\%} & $k^+<0.5$\parnotemark{c}\parnote[d]{additional surface characterization} \\
\cite{Panchanathan2018} & single & Taylor--Couette & square pillars & $4.7\e{4}$ & \SI{3}{\%} & \makecell[tl]{$k^+ = 13.6$\parnotemark{a} \\ $w^+ = 27.1$\parnotemark{a} }  \\
\hline
\cite{Du2017} & dual & channel & random & $1.2\e{5}$ & \SI{20}{\%} & not enough info given \\
\cite{Fukuda2000} & dual & \makecell[tl]{channel \\ flat plate \\ ship model} & random & \makecell[tl]{$5.0\e{4}-4.0\e{5}$ \\ $3.0\e{5}-1.7\e{7}$ \\ $9.0\e{5}-8.0\e{8}$} & \makecell[tl]{\SI{50}{\%}\parnote[e]{decreasing with \Rey} \\ \SI{50}{\%}\parnotemark{e} \\ \SI{20}{\%} } & no info given \\
\cite{Aljallis2013} & single & flat plate & random & $3.0\e{5}-3.0\e{6}$ & \SI{30}{\%}\parnote[f]{no DR for $\Rey_\text{L} > 10^6$} & no info given \\
\cite{Ling2016} & single & channel & random + ridges & $1.0\e{5}$  & \SI{36}{\%} & $k^+ = 0.68$\parnotemark{c} \\
\cite{Park2014} & single & channel & ridges & $1.0\e{5}-1.0\e{6}$ & \SI{75}{\%} & $w^+<1$\parnotemark{c} \\
\cite{Reholon2018} & single & cylindrical object & random &  $5.0\e{5} - 1.5\e{6}$ & \SI{36}{\%}\parnotemark{e} & $k^+ = 0.5$ \\
\cite{Bidkar2014} & single & channel & random & $1.0\e{6}-9.0\e{6}$ & \SI{30}{\%} & $k^+<0.5$\parnotemark{c} \\

\hline
\end{tabularx}
\parnotes
\caption{Overview of literature on drag reduction (DR) of turbulent flows over superhydrophobic (SH) surfaces, illustrating different surface design parameters $k^+$ and $w^+$ corresponding to the largest drag reduction found by different authors. The horizontal line seperates the \textit{low $\Rey$ turbulence} from the \textit{high $\Rey$ turbulence} as introduced in section 2. \label{table:DRtable}}
\end{sidewaystable}

\subsubsection*{Low $\Rey$ turbulence}
Using channel flow, \cite{Daniello2009} studied a variety of SH surfaces consisting of streamwise aligned ridges, with varying ridge spacing $w^+ = \numrange{1}{4}$. Over the whole range of $3 \times 10^3 \leq \Rey \leq 6 \times 10^3$, a DR of \SI{50}{\percent} was found~\citep{Daniello2009}. The dependence of DR on surface feature size has also been studied using Direct Numerical Simulations (DNS) by \cite{Martell2009}, finding good agreement to the work of \cite{Daniello2009}. More recent DNS of streamwise SH ridges in channel flow by \cite{Park2013}, showed a maximum DR when the ridge spacing was similar to the spacing between near-wall turbulent structures $w^+ = 100$. The work of \cite{Park2013} was able to isolate the effect of the SH surface, since it was modelled as a flat surface with an alternating no-slip and no-shear boundary condition. Effects of roughness on the flow that would play a role in experiments, either from a non-flat gas-liquid interface or from surface features that protrude through the viscous sublayer, could therefore be ruled out.

Rather than a surface of well defined feature size and geometry, a porous surface of random roughness structure was used by~\cite{Srinivasan2011}. The inner cylinder of their Taylor--Couette was was made superhydrophobic by spraycoating a mixture of PMMA fibres and low surface energy fluorodecyl POSS molecules. Nonetheless are the resulting surface roughness parameters similar to that of~\cite{Daniello2009}. From the work of \cite{Srinivasan2015} we calculate the average roughness height at the maximum $\Rey = 8\times 10^4$ to be about $k^+ = 1.5$ and the mean roughness spacing $w^+ = 2.5$. The maximum $\Rey$ also resulted in the largest DR of $\SI{22}{\percent}$. Another study in Taylor--Couette, of similar \Rey, but with much larger surface roughness parameters of $k^+ = 27$ and $w^+ = 14$ formed by a SH pillar structure, found only \SI{3}{\percent} DR~\citep{Panchanathan2018}. When instead of large SH pillars, large streamwise aligned SH ridges were used in Taylor--Couette, an optimal groove spacing of $w^+ = 35$ was found for achieving a maximum DR of $\SI{35}{\percent}$~\citep{vanBuren2017}. For the smallest groove spacing tested, $w^+ = 2$, no DR was found. The base line drag used in the definition of the drag reduction is very important. Where \cite{vanBuren2017} used their wetted surface as the baseline, was a smooth surface used for the baseline drag by~\cite{Panchanathan2018}. When the baselines are defined equally, the difference in DR found between both studies will be much smaller.

\subsubsection*{High $\Rey$ turbulence}
\cite{Ling2016} measured the velocity in the inner part of the turbulent boundary layers over SH surfaces subjected to single-phase flow. Surfaces were made SH by means of spraycoating, resulting in a random oriented roughness, and by etching and coating, giving both ridges and random oriented roughness. Measurements were done in a water tunnel, operated at $1 \times 10^5 \leq \Rey \leq 3 \times 10^5$. Their results revealed a delicate balance between the contribution of viscous stresses and Reynolds stresses to the wall shear stress. This balance determines whether DR is found (viscosity dominates), or the surface roughness increases the drag (turbulence dominates). It was found that when the roughness $k^+ \gtrapprox 1$, the Reynolds stresses become the main contributor to the wall shear stress, and less DR was found~\citep{Ling2016}.

The number of studies we found that combine a SH surface and air injection (two-phase flow) is limited. \cite{Du2017} only found DR when air was being injected through their SH surface. The DR was the result of weakened near-wall vortices, pushed away from the SH surface, and smaller shear rates on top of the SH surface~\citep{Du2017}. A variety of flow geometries was studied by~\cite{Fukuda2000}: rectangular pipe flow ($5\times 10^4 \leq \Rey \leq 4 \times 10^5$), flat plate ($3\times 10^5 \leq \Rey_\text{L} \leq 1.7 \times 10^7$), and ship models in a towing tank ($9\times 10^5 \leq \Rey_\text{L} \leq 8 \times 10^8$). For the pipe flow and flat plate experiments, the maximum DR of \SI{50}{\percent} was found to decrease with $\Rey$ to $\sim \SI{0}{\percent}$. Negligible influence of an increased air injection rate on DR was observed for all flow geometries~\citep{Fukuda2000}.

One of the few experiments in the high $\Rey$ turbulence regime that uses a surface with a geometrically well defined pattern is done by \cite{Park2014} ($1\times 10^5 \leq \Rey \leq 1 \times 10^6$), allowing for a direct comparison to the work of \cite{Daniello2009} ($3\times 10^3 \leq \Rey \leq 6 \times 10^3$) in the low $\Rey$ turbulence. Both studies made use of a fully turbulent, single-phase channel flow  over a surface of streamwise SH ridges. \cite{Daniello2009} suggested an optimum ridge spacing of $w^+ = 5$, which is equal to the size of the viscous sublayer. \cite{Park2014} however, found their maximum DR for $w^+ < 1$. This is a difference typically found between studies in the low- and the high $\Rey$ turbulence regime, as can also be seen in table~\ref{table:DRtable}.

\subsection{The air plastron}
The air layer captured between the SH surface and the water is commonly referred to as the air plastron. When the SH surface transits from a non-wetted Cassie-Baxter state to a wetted Wenzel state, the plastron and the DR are lost. Since the Wenzel state is typically the thermodynamically more favoured state, it is therefore crucial to prevent or delay this transition. This can for instance be achieved by reducing the size (diameter or $w^+$) of the asperities in which the gas is trapped to increase the Laplace pressure, or by increasing the hydrophobicity of the surface. The diffusion of gas from the plastron into the liquid is another factor to minimize in order to sustain DR, which can for instance be achieved by increasing the amount of saturated gas in the liquid~\citep{Lv2014,Xiang2017}.

In the experiments by \cite{Srinivasan2015}, the SH surface was not fully submerged, resulting in a connection between the plastron and the air present in the room. More DR (\SI{22}{\percent}) was found compared to the case where the air-layer is isolated ($\text{DR} = \SI{15}{\percent}$) for the same $\Rey$~\citep{Srinivasan2015}. When the surface is exposed to flow, the loss of plastron volume can be described by a convection-diffusion mechanism. Larger flow velocities give shorter effective diffusion lengths, resulting in an accelerated transport of gas from the plastron into the the liquid~\citep{Xiang2016}. Video recordings of the plastron exposed to turbulent flow ($5.0 \e{5} \leq \Rey_L \leq 1.5\e{6}$) showed constant movement and variations in the thickness of the plastron, caused by pressure fluctuations in the turbulent boundary layer~\citep{Reholon2018}. \cite{Du2017} found DR when injecting air through a pinhole in their SH surface. The amount of injected air was not enough to form an air bubbly flow, but enough to maintain a plastron that was thick enough to prevent the surface roughness features from contacting the liquid. When the air injection was stopped, the air plastron became thinner, and roughness effects started to play a role~\citep{Du2017}. When the roughness elements are exposed to the flow, the Reynolds stresses become the main contributor to the wall shear stress, resulting in less DR~\citep{Ling2016}.

\add{For this particular reason,~\citet{Gose2018} suggested to measure the surface characteristic contact angle hysteresis under higher than ambient pressures.} Also mechanical interactions between the plastron and solid pollutants in the liquid phase can decrease the plastron stability. Collisions between particles added to the flow and the plastron were shown to shorten its lifetime by about $\SI{50}{\%}$~\citep{Hokmabad2017}. Once the air plastron is destroyed and the surface has transited to the wetted state, energy is required to reverse the surface to the un-wetted state. Different studies explored for instance film boiling, water splitting by electrolysis and the injection of air bubbles into the boundary layer (dual-phase flow) to achieve this~\citep{Saranadhi2016,Panchanathan2018}.

\subsection{Bubbly drag reduction}
\add{The introduction of air bubbles to a flow can also result in reduced skin friction drag.} A typical approach is to inject air bubbles close to (or in) the boundary layer.
We refer to the review articles by~\cite{Ceccio2010} and \cite{Murai2014} for an overview of various studies on bubbly DR.
Early air-lubrication DR experiments, in which gas micro-bubbles were injected (or created) in the (turbulent) boundary layer, showed an increase in DR with increasing air injection rate, and a decrease in DR with increasing Reynolds number~\citep{McCormick1973,Madavan1985,Watanabe1998}. Up to \SI{80}{\percent} reduction of skin friction drag using microbubble injection was reported~\citep{Madavan1984}. This DR was attributed to a thickening of the viscous sublayer (so a smaller gradient in the velocity profile near the wall) caused by the microbubbles that were present in the near-wall buffer layer~\citep{Ceccio2010}.

For Taylor--Couette flow, for low $\Rey$ and microbubble injection, the drag reduction was shown to be due to the weakening or even destruction of the Taylor-rolls, due to the rising microbubbles. This gravity effect (controlled by the Froude number) indeed decreases with increasing Reynolds numbers~\citep{Sugiyama2008,Lohse2018}. More recent research showed the influence of the bubble size on DR, concluding that the existence of large, deformable bubbles, i.e. those that have a large Weber number, is crucial for drag reduction in high $\Rey$ turbulent flows~\citep{Lu2005,vandenBerg2005,vanGils2013,Verschoof2016,Spandan2018}. In these papers, the increase of the Weber number with increasing $\Rey$ is used to explain the enhanced bubbly DR that is typically found for larger $\Rey$~\citep{vandenBerg2005}.

Although the principle of air bubbly DR is not yet fully understood, it is clear that the effect is largest when the bubbles are close to, or in, the boundary layer. For flat plate experiments, the skin friction bubbly drag reduction is commonly limited to the first few metres downstream of the air injector~\citep{Watanabe1998,Sanders2006}. Further downstream, turbulent diffusion causes bubbles to move away from the wall~\citep{Murai2014}. A similar mechanism was observed in Taylor--Couette flow, where strong secondary flows transport bubbles away from the inner cylinder, resulting in a decrease of DR~\citep{vandenBerg2007,Fokoua2015,Verschoof2018}.  \\

\subsection{Roughness}
To create a SH surface, some form of roughness has to be introduced on the surface, to facilitate an asperity where air can be trapped. Since any form of surface roughness increases the drag on a wall-bound flow, we therefore deal with opposing effects in drag reduction using SH surfaces: drag reduction due to air (plastron) lubrication and drag increase from the added roughness. We refer to the reviews of \cite{Jimenez2004} and \cite{Flack2010} for a comprehensive overview of studies towards the influence of roughness on turbulent flows.

Three different roughness regimes are distinguished. In the hydrodynamically smooth regime, when the equivalent sand roughness is less than the thickness of the viscous sublayer ($k^+ < 5$), the surface can be regarded as smooth~\citep{Schlichting}. The perturbations in the flow that are generated by the roughness features of the surface are completely damped out by the viscosity~\citep{Flack2014}. When the roughness $k^+$ increases, parts of it will extend through the viscous sublayer, corresponding to the transitionally rough regime ($5 \leq k^+ < 70$). The log-law that describes the velocity profile close to the wall shifts inwards, maintaining its shape, but reduced in magnitude. The mean velocity profile in the bulk of the flow however stays unaffected by the roughness~\citep{Flack2014}. Hence, universality is only seen for the larger length scales of the flows~\citep{Pope}.

The wall shear stress in the transitionally rough regime is composed of a combination of viscosity and pressure drag on the roughness elements. With increasing roughness height, the contribution of pressure drag increases~\citep{Verschoof2018EPJE}. In the fully rough regime ($k^+ \geq 70$), the pressure drag heavily dominates over viscosity. As a result, the shift in the log law (the roughness function $\Delta U^+$), scales linearly with $k^+$, and the skin-friction coefficient becomes independent of $\Rey$~\citep{Flack2014}.

Although the size of the roughness $k^+$ gives a good indication for the state of roughness: hydrodynamically smooth, transitionally rough or fully rough, which also depends on the geometry of the roughness. For instance, a stepwise geometry that consists of steep slopes will transit to the fully rough regime at smaller $k^+$ than a roughness of more gentle slopes~\citep{Busse2017}. Similarly, a surface of very closely packed roughness elements (high solidity), or a surface where the roughness elements are sparse (high porosity) will behave more like a surface of smaller $k^+$~\citep{MacDonald2016}.
In the context of Taylor--Couette turbulence, roughness effects were analyzed by \cite{Zhu2018} and \cite{Berghout2018}, who found the same universal $\Delta U^+ \left( k^+ \right)$ for the velocity reduction as was found by \cite{Nikuradse1933} for pipe flow.

%% file: experimental.tex
% !TEX root = Main document.tex
\section{Experimental method}

\subsection{Experimental setup}
\begin{figure}
    \centering
   \includegraphics{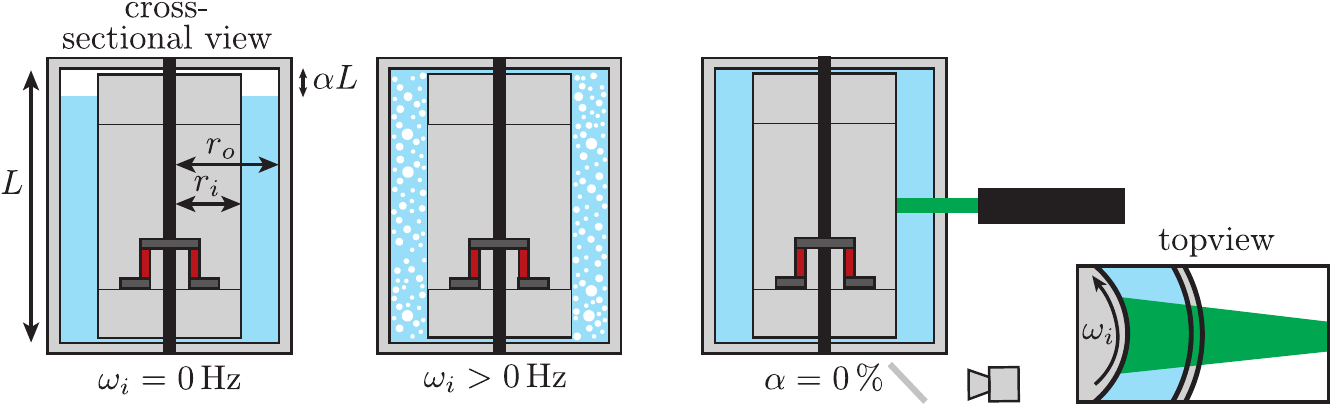}
    \caption{Schematic overview of the measurement setup. Shown are the outer and inner cylinder, of which the latter consists of three sections. The middle section is connected to the driving shaft by means of a torque sensor, which is also shown in the figure. The gap between the two cylinders $r_o - r_i$ is filled with water and air, of which the quantity of the air is expressed by means of a void fraction $\alpha$, ranging between \SI{0}{\percent} and \SI{6}{\percent}. When the inner cylinder is rotating ($\omega_i > \SI{0}{\hertz})$, bubbles are formed and distributed in radial and axial direction over the gap due to turbulent mixing. Particle Image Velocimetry (PIV) measurements can be done only when there are no bubbles present in the working liquid ($\alpha = \SI{0}{\percent}$). The PIV lasersheet is placed at cylinder mid-height and the flow is observed through a window in the bottom-plate using a mirror and a camera.}\label{fig:T3C}
\end{figure}

All experiments were performed in the Twente Turbulent Taylor-Couette (T$^3$C) facility described in~\citet{VanGils2011} and shown in figure~\ref{fig:T3C}.
 It consists of two independently rotating concentric cylinders of length $L = \SI{0.927}{\metre}$. The inner cylinder is fabricated from grade 316 hydrophilic stainless steel. The outer cylinder is cast from clear acrylic, which allows for full optical access to the flow between the cylinders. The inner radius of the outer cylinder is $r_o = \SI{0.279}{\metre}$ %0.2794
and the outer radius of the inner cylinder equals $r_i~=~\SI{0.200}{\metre}$, thus the radius ratio is $\eta = r_i / r_o = 0.716$. %0.7158
 The resulting gap has a width $d = r_o - r_i =
\SI{0.079}{\metre}$
 and was filled with \add{fully air saturated deionized} water. For inner cylinder rotation, the Reynolds defined, based on the gap width and velocity of the inner cylinder is
\begin{equation}\label{eq:Rey}
\Rey = \frac{\omega_i r_i d}{\nu},
\end{equation}
where $\omega_i$ is the angular velocity of the inner cylinder and $\nu$ is the kinematic viscosity of the working fluids. The inner cylinder rotates at frequencies in the range $\omega_i = \SIrange{5}{18}{\hertz}$, while the outer cylinder is kept stationary. Typical values used in this research range from $\Rey = 5 \times 10^5$ to $\Rey = 1.8 \times 10^6$. The system is actively cooled to keep the temperature of the working fluid at $\SI{21}{\degreeCelsius} \pm \SI{0.5}{\degreeCelsius}$.

By partly filling the apparatus, as in figure~\ref{fig:T3C}, we vary the volume fraction of air $\alpha$ in the working fluid from \SI{0}{\percent}~to~\SI{~6}{\percent}. The turbulence mixes the air and water, generating bubbles that are distributed over the height and over the gap between cylinders~\citep{vanGils2013}.

\subsection{Torque measurements}
The inner cylinder is composed of three sections. The torque exerted by the fluid on the inner cylinder is measured using a Honeywell 2404-1K hollow reaction torque sensor that is placed inside the middle section of the inner cylinder as indicated in figure~\ref{fig:T3C}. Only the torque on the middle section of length $L_{\text{mid}} = 0.536$~m is taken into account to reduce end-plate effects between the rotating lid of the inner cylinder and the stationary lid of the outer cylinder. We express the torque in non-dimensional form using the skin friction coefficient:
\begin{equation}\label{eq:cf}
C_f = \frac{\mathcal{T}}{L_{\text{mid}}\rho\nu^2\Rey^2}
\end{equation}
where $\mathcal{T}$ denotes the torque, $\rho$ and $\nu$ are the density and kinematic viscosity, respectively, of the working fluid. \\
The drag reduction for the {hydrophobic} coating and the hydrophilic reference is determined using equation~(\ref{eq:DR1}).
\begin{equation}\label{eq:DR1}
\text{DR}(\alpha) = 1 - \frac{C_{f}(\alpha)} {C_{f}(\alpha=0)}
\end{equation}
This shows the influence of adding bubbles to the flow on the drag. The difference in drag reduction $\Delta\text{DR}$ between a {hydrophobic} inner cylinder (IC) and a hydrophilic IC is defined as
\begin{equation}\label{eq:deltaDR}
\Delta\text{DR}(\alpha) = DR_\text{hydrophobic}(\alpha) - DR_\text{hydrophilic}(\alpha)
\end{equation}
In order to provide insight into the influence of the {hydrophobic} IC on the flow, we define a {\it net} drag reduction as
\begin{equation}\label{eq:DR2}
\text{DR}_\text{\it net}(\alpha) = 1 - \frac{C_{f,\text{hydrophobic}}(\alpha)}{C_{f,\text{hydrophilic}}(\alpha = 0)}
\end{equation}
Here $C_{f,\text{hydrophobic}}(\alpha)$ is the skin friction coefficient for the {hydrophobic} IC and different values of $\alpha$, and $C_{f,\text{hydrophilic}}(\alpha = 0)$ is the skin friction coefficient for the smooth hydrophillic inner cylinder, without air bubbles present in the flow ($\alpha$ = 0). In equation~(\ref{eq:DR1}) the focus is only on the influence of bubbles on the drag, with either a {hydrophobic} or a hydrophilic IC, whereas equation~(\ref{eq:DR2}) show the influence of both drag reducing measures: bubbles and a {hydrophobic} coating.

\subsection{{hydrophobic} coating}
In the {hydrophobic} case, the IC of the T$^3$C is fully coated with a 3M~Membrana Accurel\textsuperscript{\textregistered}~PP~2E~HF flatsheet membrane. This porous hydrophobic polypropylene material is commercially available in the large quantities that are needed to cover the complete IC.
This coating is supplied on rolls that have a width of about \SI{27}{\cm}. It is attached to the inner cylinder using double-sided adhesive tape. From visual inspection it is estimated that \SI{99}{\percent} of the IC is covered by the coating, see figure~\ref{fig:photo_coating_glow}.

We used a Dataphysics OCA 15EC device to measure the contact angle hysteresis for both the {hydrophobic} coating and the reference case, which is the hydrophilic, uncoated steel IC. The {hydrophobic} coating has advancing and receding water contact angles of $\SI{152}{\degree} \pm \SI{2}{\degree}$ and $\SI{120}{\degree} \pm \SI{5}{\degree}$, respectively. For the hydrophilic IC, an advancing contact angle of $\SI{93}{\degree} \pm \SI{2}{\degree}$ was found. The receding contact angle is set at \SI{10}{\degree}, which is the lowest angle the setup could measure.

\begin{figure}
\centering
\includegraphics[width=\textwidth]{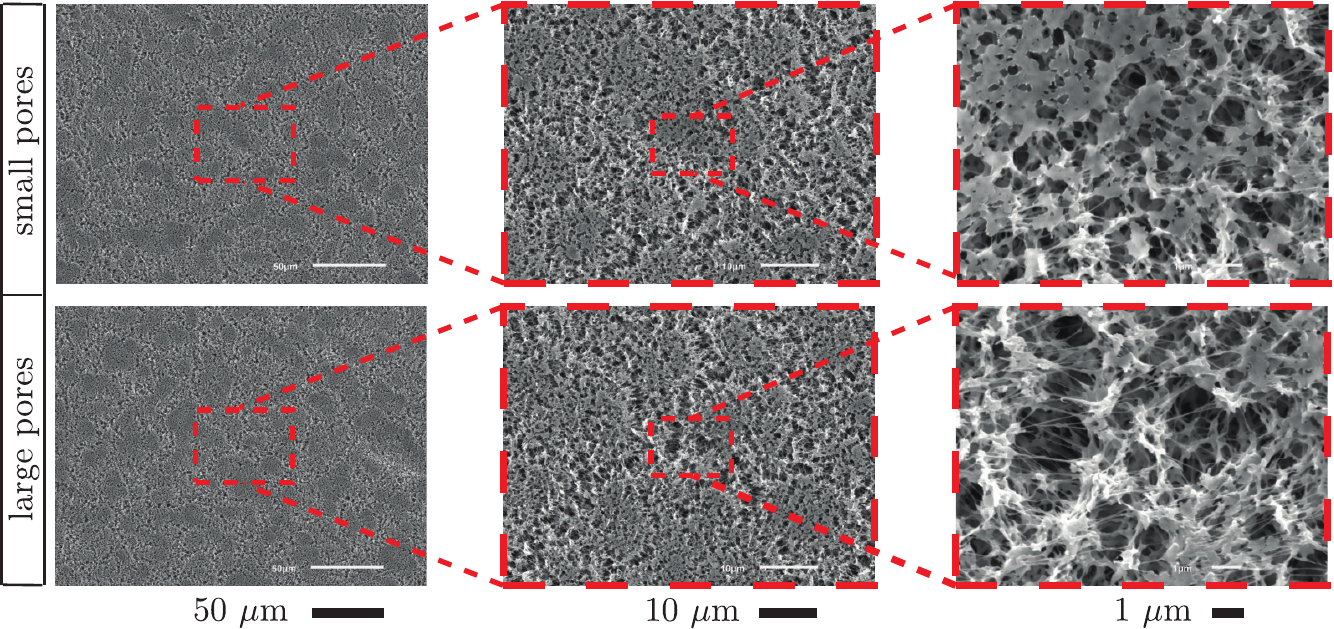}
\caption{SEM photos of the side of the coating that is exposed to the flow. The top row focuses on a region composed of smaller pores. The bottom row shows a region with larger pores.}\label{fig:SEM-bottom}
\end{figure}

\subsection{Roughness}
The machining process to fabricate the standard hydrophilic IC gives a surface roughness of $k_{ic} = \SI{1.6}{\um}$. The viscous length scale $\delta_\nu$ is derived from the measured torque data, as discussed in~\citep{hui13}. For the maximum Reynolds number $\Rey_\text{max} = 1.8 \times 10^6$ used in this research, the viscous length scale reaches its lowest value of $\delta_\nu = \SI{1.9}{\um}$. The resulting roughness in wall units $k^+_{ic} \approx 0.8$. Therefore, the uncoated hydrophilic IC can be assumed to be a hydrodynamically smooth surface.

\begin{figure}
\centering
\includegraphics{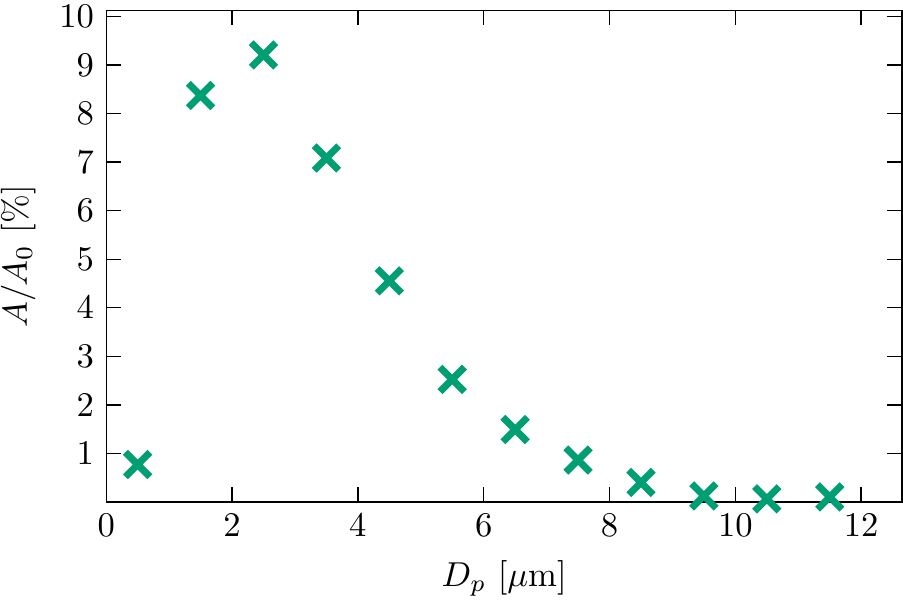}
\caption{Pore size diameter $D_p$ (roughness) distribution of the coating as fraction of coverage $A$ of total area of the coating $A_0$. A range of length scales is observed, corresponding to the different regions identified in figure~\ref{fig:SEM-bottom}. The equivalent circle diameter has been used as a measure of the size $D_p = 2 \sqrt{A_p/\pi}$.} \label{fig:poresize_plot}
\end{figure}

The average roughness of the coating is analysed from its pore size using 24 different SEM images, made using three different magnifications, as shown in figure~\ref{fig:SEM-bottom}. \add{The coating consists of an isotropic sponge-like structure, meaning that the cross section looks similar to the top and bottom surface. Therefore we use SEM images of the top surface to evaluate the size and roughness of the pores.} The SEM images show a distribution of pore sizes, in the range of \SIrange{1}{10}{\um}. The pores that correspond to the smaller length scale are found in regions separated by pores of the larger length scale. The size distribution is quantified with the image processing program ImageJ, using edge detection of the thresholded image (Analyse Particles tool). Eight different images of the smallest magnification were used for this. The images were pre-processed by subtracting a sliding background and by applying a local mean threshold algorithm. Eroding and dilation was used to remove small scale noise. It is difficult to define the error for the pore size distribution, since it is difficult to evaluate the edge of a pore from the perspective of the flow. For instance in figure~\ref{fig:SEM-bottom}, when inspecting the large pores at highest magnification, we see thin thread-like fibres that span across a pore. Whereas in the image analysis this might be detected as an edge, the flow might experience this differently. However, from the SEM images and the pore size distribution it is clear that multiple roughness length scales are present on the surface of the coating. Hence, dependent of $\Rey$, a larger or smaller fraction of the surface plays a role in influencing the flow. The resulting distribution of binned pore sizes is shown in figure~\ref{fig:poresize_plot}. The combined area of all pores with diameter $D_p$ over the total area, the fraction $A/A_0$, is plotted versus $D_p$. A maximum is found at $D_p = \SI{2.5}{\um}$, although the larger length scales that are more relevant to the flow are also found.

\subsection{Experimental procedure}
During a measurement period of one hour, the IC was accelerated in steps from \SI{5}{\hertz} to \SI{18}{\hertz}, corresponding to a range of $\Rey$ between $5\times10^5$ and $1.8\times10^6$, while continuously measuring torque exerted by the fluid on the inner cylinder.
Every variation of a {hydrophobic} IC with $\alpha > \SI{0}{\%}$ was measured four times. Between changing the volume percent of air $\alpha$, the reference case of $\alpha = \SI{0}{\%}$ air is measured twice, to account for changes to the coating caused by the flow itself. An overview of the measurements is shown in order of execution in table~\ref{table:measurement_order}.
\begin{table}
\centering
\begin{tabular}{ | l | l | l | c | }
\cline{1-4}  \\[-0.9em]
 Surface & $\alpha$ [\%]  & $\omega_i$ [Hz] & Measurements \\ \cline{1-4} \\[-0.9em]
hydrophobic &  0  & 	\numrange{5}{18} & 2 \\
					  &  2 	&  	5--18 & 4 \\
					  &  0 	& 		5--18 & 2 \\
					  &  4 	& 		5--18 & 4 \\
					  &  0 	& 		5--18 & 2 \\
					  &  6 	& 		5--14 & 4\\
					  &  0 	& 		5--18 & 2 \\
\cline{1-4} \\[-0.9em]
hydrophilic   & 0 	& 		5--18 & 2 \\
					  & 2 	& 		5--18 & 3\\
					  & 4 	& 		5--13.4 & 3\\
  					  & 4 	& 		15.8--18  & 3\\
					  & 6 	& 		5--14 & 3 \\ \cline{1-4}

\end{tabular}\caption{Overview of the measurement parameter space, in order of execution. Between changing the volume percentage of air $\alpha$, the reference case of $\alpha = \SI{0}{\%}$ air was measured twice, to account for changes to the coating caused by the flow itself. Deviations from the standard frequency range $\omega_i = \SIrange{5}{18}{\hertz}$ were the result of heavy vibrations in the system, forcing us to skip a certain frequency range.} \label{table:measurement_order}
\end{table}

\subsection{Flow visualisation}
\begin{figure}
\centering
\includegraphics[width=0.4 \textwidth]{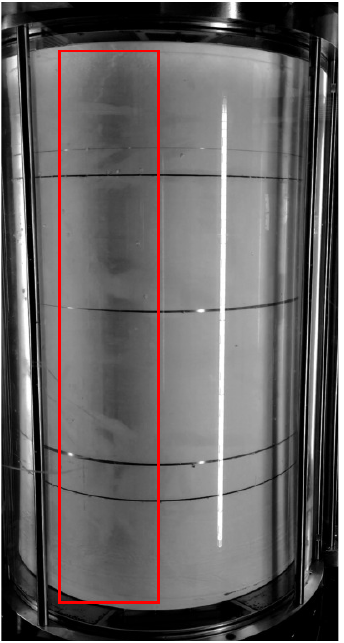}
\caption{Digitally enhanced photograph of the inner cylinder, covered with the {hydrophobic} coating and visible through the transparent outer cylinder. Estimated is that \SI{99}{\percent} of the inner cylinder is covered by the {hydrophobic} coating. The silvery reflection that is typically associated with the presence of an air plastron, is visible as a darker shaded region. This plastron can only be observed under certain angles of incident light. The curved surface of the inner cylinder explains why the plastron is only visible in a narrow vertical band.}\label{fig:photo_coating_glow}
\end{figure}

A Nikon D800E camera was used to capture still images of the flow. This provides insight on the presence of an air plastron: air captured by the coating is visible in the form of a silvery reflection on the surface \citet{Shirtcliffe2006,McHale2009,Daniello2009,Poetes2010,Mchale2011,Dong2013,Park2014,Park2015,
Saranadhi2016}. This can be seen in figure~\ref{fig:photo_coating_glow}, where the highlighted area points out locations where the incident light is of the right angle to see the plastron. It was found that an air plastron was present during the measurements featuring a {hydrophobic} IC.
\add{To test the stability of the plastron and force the surface in a wetted state, the surface tension was lowered by adding TritonX surfactant whilst rotating the inner cylinder at $\Rey = 1.0\e{6}$ with single phase flow conditions. From image analysis it was determined that only after \SI{1.5e-4}{\mol \per \litre} TritonX was added, the silvery reflection had completely dissapeared. At this concentration of TritonX the surface tension is about \SI{35}{\milli \newton \per \metre}~\citep{Gobel1997}. So both surface tension and Laplace pressure are about half the value of that for pure water.}

\subsubsection{Velocity profile measurements}
Particle Image Velocimetry (PIV) was used to obtain local flow field information. We measured the velocity field in the $(r,\theta)$ plane; $u_\theta = u_\theta (r,\theta,t)$ and $u_r = u_r (r,\theta,t)$. This can only be achieved for single-phase flow with $\alpha = 0$, since the air bubbles otherwise scatter the light significantly. The laser light sheet (Quantel Evergreen 145 laser, \SI{532}{\nm}) used to illuminate the seeding particles added to the flow (Dantec fluorescent polyamide, with a distribution of diameters $\leq \SI{20}{\um}$) was placed at mid-height of the cylinder. Images were captured using a LaVision sCMOS ($2560\times2160$ pixel) camera through the window in the bottom plate of the setup. Figure~\ref{fig:T3C} gives a schematic overview of the measurement setup. Average velocity fields were calculated from 1000 image pairs using LaVision DaVis software in a multi-pass method, starting at a window size of $64\times64$ pixel decreasing to a final size of $24\times24$ pixel with \SI{50}{\percent} overlap.
A calibration is required to transform pixels to meters. To this end image analysis is used to locate the edges of the inner and outer cylinder. Since the measured fields are in Cartesian coordinates, a coordinate transformation is necessary to obtain finally the radial and azimuthal velocities $u_r$ and $u_\theta$ in the cylindrical coordinate system.

%% file: results.tex
% !TEX root = Main document.tex

 \section{Results and discussion}
\subsection{Torque measurements}
First, the results of the torque measurements single-phase flow are presented and discussed, where drag reduction (DR) is purely the result of a {hydrophobic} surface capturing an air plastron. Second, we show the results for two-phase flow, where air bubbles are added to the flow, that provide bubbly DR and might also add to the stability of the air plastron.
\subsubsection*{Single-phase flow}
\begin{figure}
\centering \includegraphics[width=0.76\textwidth]{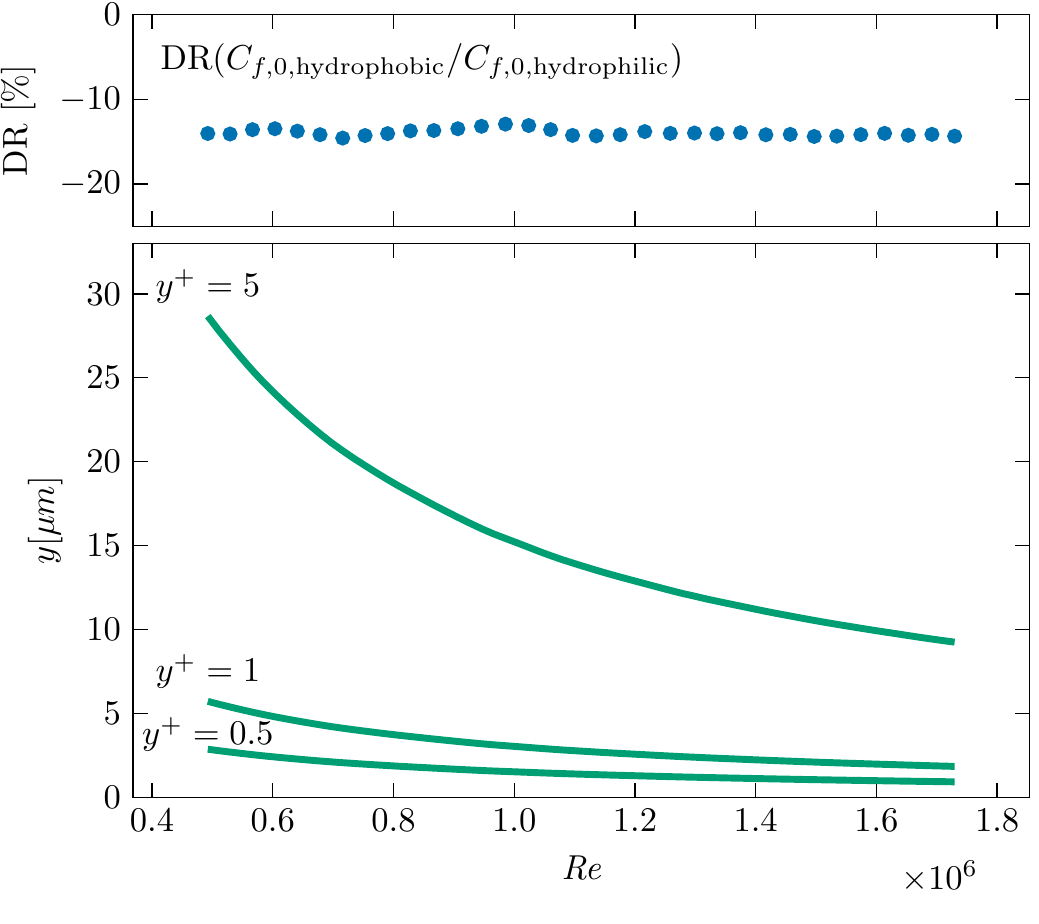}
\caption{Top figure: plot of the drag reduction (DR) as in equation~(\ref{eq:DR2}) versus $\Rey$, with $\alpha = 0$ for the {hydrophobic} inner cylinder. Bottom figure: evolution of the thickness of the viscous sublayer ($y^+ = 5$), the viscous length scale ($y^+ = 1$) and half the viscous length scale ($y^+ = 0.5$) with $\Rey$. The design parameters $w^+ < 1$ and $k^+ < 0.5$ for the {hydrophobic} surface are suggested by \cite{Park2014} and \citep{Bidkar2014} respectively to result in drag reducing behaviour of the surface and are shown here as a reference for the reader. These values are derived from the torque measurements and give therefore an averaged, global value. From figure~\ref{fig:3Re} we find that for our lowest $\Rey$ tested, the majority of the roughness length scales is below $k^+ = 1$ and part of it is below $k^+ = 0.5$. The DR plot however shows a nearly constant increase of the drag by about $\SI{14}{\%}$ over the whole range of $\Rey$ measured.} \label{fig:DR_delta_nu_Dp_groupplot}
\end{figure}

\begin{figure}
\centering \includegraphics[width=0.8\textwidth]{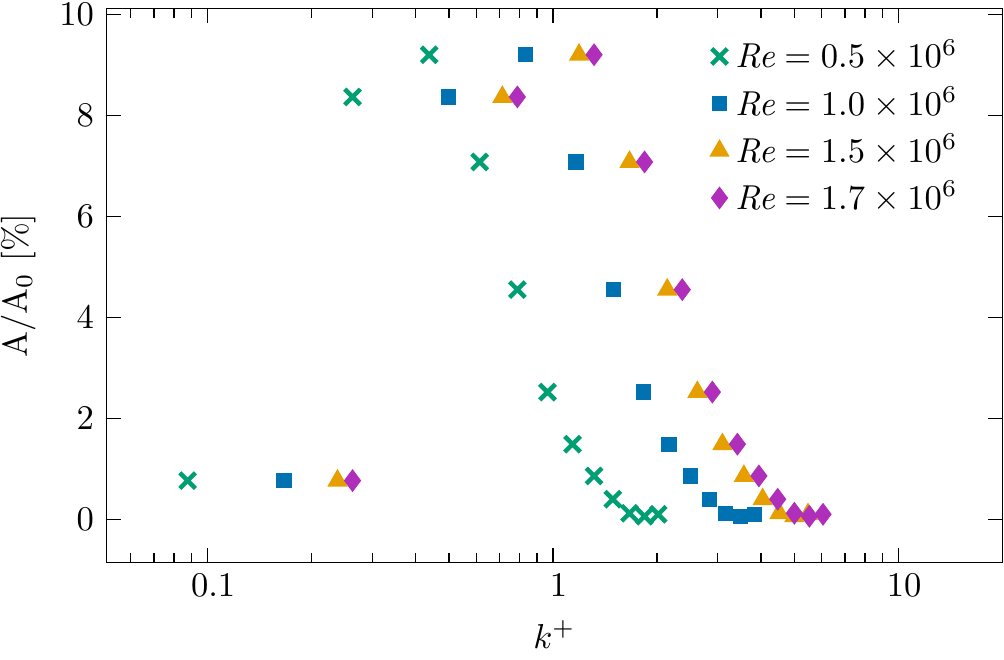}
\caption{Roughness distribution of the surface as coverage fraction of the total area $A/A_0$, expressed in wall units for four different values of $\Rey$. Apart from the maximum and the minimum value of $\Rey$ used in this research, the normalized roughness for two intermediate values of $\Rey$ are shown as well. The wall unit normalization is obtained using data from the torque measurements that gives an averaged, global value of the wall shear stress.} \label{fig:3Re}
\end{figure}

\noindent
In the top of figure~\ref{fig:DR_delta_nu_Dp_groupplot}, the drag reduction (DR) as defined in equation~(\ref{eq:DR2}) is plotted versus $\Rey$, for the {hydrophobic} inner cylinder (IC) with $\alpha = 0$. This shows an increase in the drag of about \perc{14} over the whole range of $\Rey$ measured. The bottom figure shows the evolution of the thickness of the viscous sublayer ($y^+ = 5$), the viscous length scale and the design parameter suggested by \cite{Park2014} ($y^+ = 1$), and the design parameter suggested by \cite{Bidkar2014} ($y^+ = 0.5$), with the Reynolds number. In figure~\ref{fig:3Re} we show the roughness of the surface expressed in wall units for four different values of $\Rey$: the minimum, the maximum and two intermediate values, for which we use the same data as in figure~\ref{fig:poresize_plot}. For the lowest $\Rey = 0.5 \times 10^6$ , the majority of the roughness length scales is less than $k^+ = 1$ and part of it is even less than $k^+ = 0.5$. However, over the whole range of $\Rey$ measured we find a constant increase of the drag by about $\SI{14}{\%}$. For a rough, wetted surface, an increase of drag with $\Rey$ is expected~\citep{Flack2010}. However, we do not observe wetting of the surface, which would manifest itself through disappearance of the silvery reflection that indicates the presence of an air plastron. Hence, we assume the surface to maintain its Cassie-Baxter state throughout the course of the experiments. \add{Nonetheless, even for a wetted surface, an increase in $C_f$ is very surprising, given the average value of $k^+ < 5$, which would indicate a hydrodynamically smooth surface following~\citep{Schlichting}. A similar result of drag increase with $k^+ < 5$ was also reported by~\cite{Gose2018}. Contrary to the results observed of these authors of increasing DR with decreasing $k^+$ for the same surface, we find a more or less increased constant drag over the whole range of our values of $\Rey$ and hence $k^+$. Our results might be explained by following the analysis of~\cite{Gose2018}, who state that the value of the $k^+$ roughness alone is not sufficient to predict the DR of a superhydrophic surface.} The work of \cite{Reholon2018} showed less DR at larger $\Rey$ due to a thinner and fragmented plastron. However, we find the DR to be more or less constant at $\SI{-14}{\percent}$. \add{When the surface is forced into a wetted state by lowering the surface tension from the addition of TritonX, no significant difference in drag is found. Obviously, the surface roughness strongly influences the drag even in the non-wetted state.}

\subsubsection*{Two-phase flow}
\begin{figure}
\centering \includegraphics{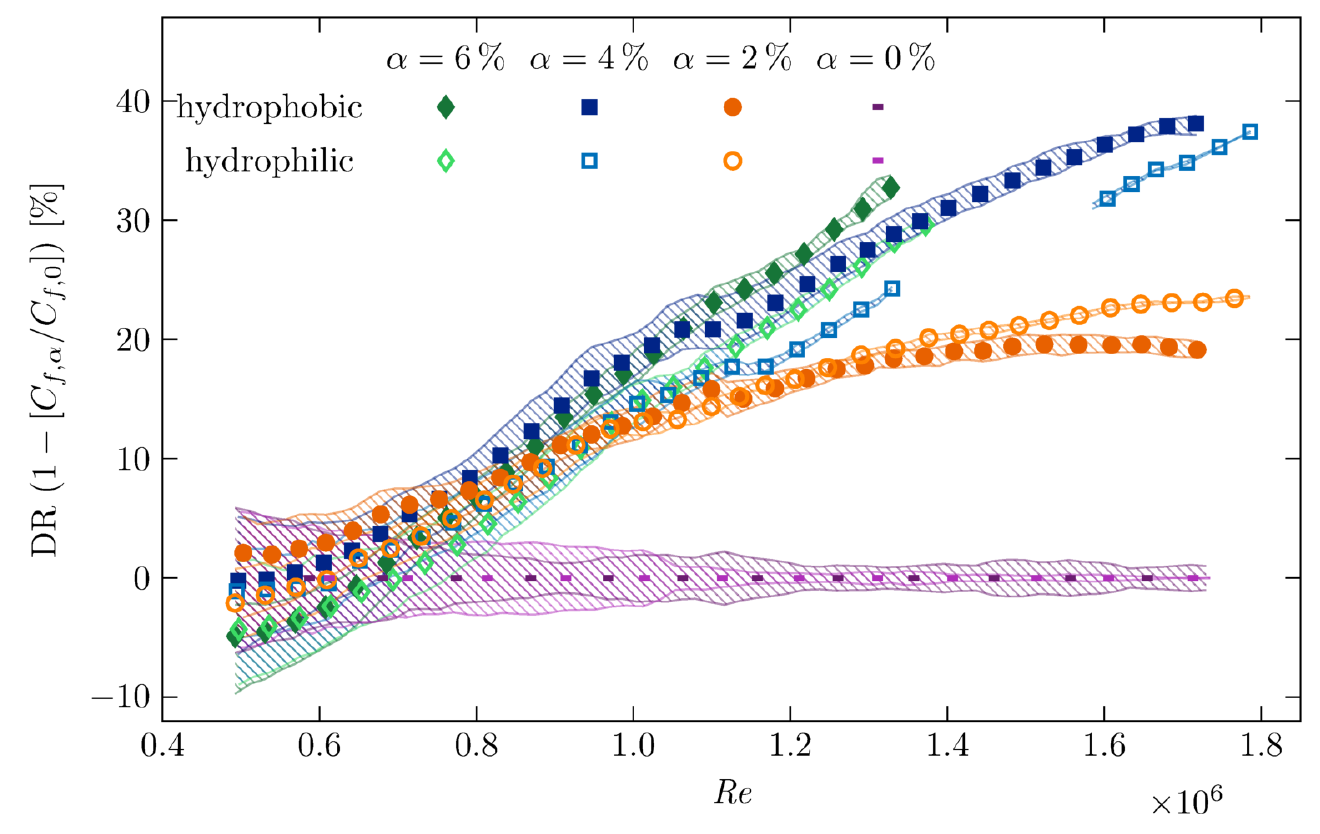}
\caption{Plot of the drag reduction (DR) based on skin friction coefficient $C_f$ as defined in equation~(\ref{eq:DR1}) versus $\Rey$. Compared to figure~\ref{fig:DR2}, is the drag reduction here determined using the same IC as used for the $\alpha > 0$ measurement (so either hydrophobic or hydrophilic) with $\alpha = 0$ ($C_{f,0}$), whereas in figure~\ref{fig:DR2} the reference is the hydrophilic IC with $\alpha = 0$. A more efficient bubbly drag reduction mechanism is found for the hydrophobic coating when the void fraction $\alpha \geq \perc{4}$. For $\alpha  = \perc{2}$ roughness effects dominate, resulting in less overall DR. Only every second data point is shown to improve readability of the plot. The shaded regions represent the spread in data. \add{The size of the error bars based on the accuracy of the torque sensor is smaller than the marker size.} Due to heavy vibrations in the setup resulting from a non-symmetric distribution of air, no data was acquired in the region between $\Rey = 1.3\times10^6$ and $\Rey = 1.6\times10^6$ for $\alpha=\perc{4}$ and for $\Rey > 1.4\times10^6$ when $\alpha=\perc{6}$.}\label{fig:DR1}
\end{figure}

\noindent
In figure~\ref{fig:DR1}, DR is shown as defined in equation~(\ref{eq:DR1}), comparing results for a \textit{hydrophobic} inner cylinder to a \textit{hydrophilic} inner cylinder, plotted versus $\Rey$. For both hydrophobic and hydrophilic cases, $\alpha = \SI{0}{\percent}$ is the reference case for determining the level of drag reduction that results from the introduction of bubbles ($\alpha > \perc{0}$) to the flow. As an additional effect, the introduction of air bubbles to the working liquid might also lead to enhanced stability of the air plastron~\cite{Lv2014}. With increasing $\Rey$, more DR is found for all measurements. This is in line with the previous findings~\citep{vandenBerg2005,vanGils2013,Spandan2018}. %In in our experiments, $\text{We} = 1$ is reached $\Rey \approx 6 \times 10^5$ and increases with Reynolds number~\citep{vandenBerg2005}.}

Comparing the {hydrophobic} IC to the hydrophilic IC, more bubbly DR is found (figure~\ref{fig:DR1}) over nearly the whole range of $\Rey$  when $\alpha \geq \perc{4}$. Only when $\Rey < 7 \times 10^5$, a slight increase of drag is found for $\alpha = \perc{6}$. In this range of low $\Rey$, the uncertainty of the torque sensor is the highest, as can be seen from the shaded areas in figure~\ref{fig:DR1} that give an indication of the repeatability of the experiments by showing the spread in the data by comparing the extremes of individual measurements. For the smallest amount of air added, $\alpha = \perc{2}$, the {hydrophobic} IC gives more DR compared to the hydrophilic IC up to $\Rey = 10^6$. For larger $\Rey$, between $1.0\times10^6$ and $1.8\times 10^6$, the {hydrophobic} IC gives less DR compared to the hydrophilic IC.

We explain the difference between $\alpha = \perc{2}$ and $\alpha \geq \perc{4}$ in figure~\ref{fig:DR1} as a result of two competing effects: (i) a more effective bubbly drag reduction in the presence of a {hydrophobic} wall, and (ii) a drag increase due to the roughness of this same {hydrophobic} wall. When the thickness of the viscous sublayer $y_\text{vsl} = 5y^+$ decreases with increasing $\Rey$, a larger fraction of the pores on the surface of the {hydrophobic} IC are of a length scale relevant to the flow, as can be seen in figure~\ref{fig:DR_delta_nu_Dp_groupplot}. The balance between these two competing effects determines whether the bubbly DR is more effective when using a rough {hydrophobic} IC compared to a smooth hydrophilic IC. For larger void fractions $\alpha \geq \perc{4}$, the more effective bubbly DR dominates, resulting in more overall DR.

\begin{figure}
\centering
\includegraphics{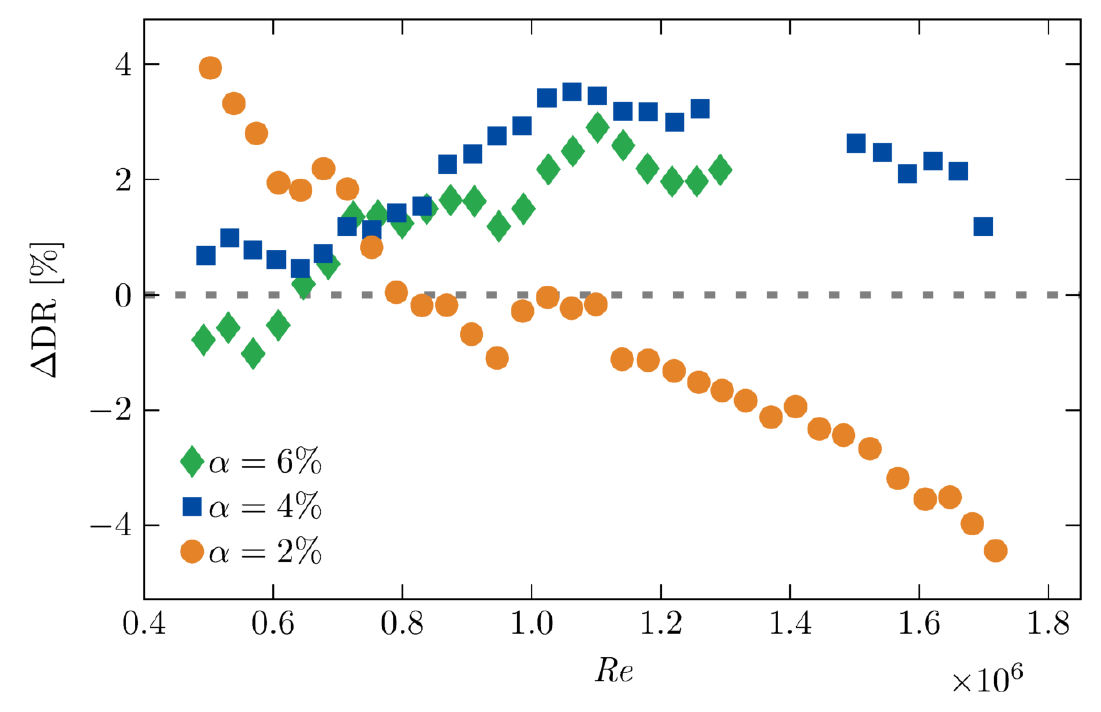}
\caption{Plot of $ \Delta\text{DR} = \text{DR}_\text{hydrophobic} - \text{DR}_\text{hydrophilic}$ from figure~\ref{fig:DR1} versus $\Rey$. For $\alpha = 2$, $\Delta \text{DR}$ decreasing with increasing $\Rey$, owing to the influence of roughness. For $\alpha \geq 4$, $\Delta \text{DR}$ shows an increase with $\Rey$, meaning that the effect of increase in bubbly drag reduction due the hydrophobic coating is stronger than the effect of roughness.}\label{fig:deltaDR}
\end{figure}

Figure~\ref{fig:deltaDR} shows the difference $\Delta \text{DR}$ between \textit{hydrophobic} and \textit{hydrophilic} from figure~\ref{fig:DR1}, as defined in equation~(\ref{eq:deltaDR}). The difference in $\Delta \text{DR}$ between $\alpha = \perc{4}$ and $\alpha = \perc{6}$ in figure~\ref{fig:deltaDR} is small compared to the difference between $\alpha = \perc{2}$ and $\alpha = \perc{4}$, or $\alpha = \perc{2}$ and $\alpha = \perc{6}$. This suggests that the influence of the {hydrophobic} IC on the bubbly DR is limited, i.e. that a minimum amount of air is required to effectively reduce the drag --- here $\alpha = \perc{4}$ --- but further increasing $\alpha$ will not result in an even stronger influence of the {hydrophobic} IC on the bubbly DR. Nor does $\Delta \text{DR}$ increase with $\Rey$ for $\Rey > 10^6$ and $\alpha \geq \perc{4}$, but rather levels off and starts to decrease. This indicates that of the two effects, the more effective DR from the {hydrophobic} wall initially grows faster with $\Rey$, until $\Delta \text{DR}$ is maximum. With further increasing $\Rey$, around $\Rey = 1.1 \times 10^6$, the balance starts to tilt and the drag increase from the roughness is now the faster growing effect, indicated by the negative slope of $\Delta \text{DR}$ for $\alpha = \perc{4}$ and $\alpha = \perc{6}$ in figure~\ref{fig:deltaDR}. It is tempting to attribute this to wetting of the coating, caused by the wall shear stresses that become larger with $\Rey$. However, if this were the case, a clear difference would be visible between the initial, and its repeated measurements for the same $\alpha$, since the wetted state is the energetically more stable one. It is safe to assume that if the coating -- or part of it -- is wetted, it will not transit back to a non-wetted state between measurements. The reason for the change in balance should therefore be sought elsewhere.

\begin{figure}
\centering \includegraphics{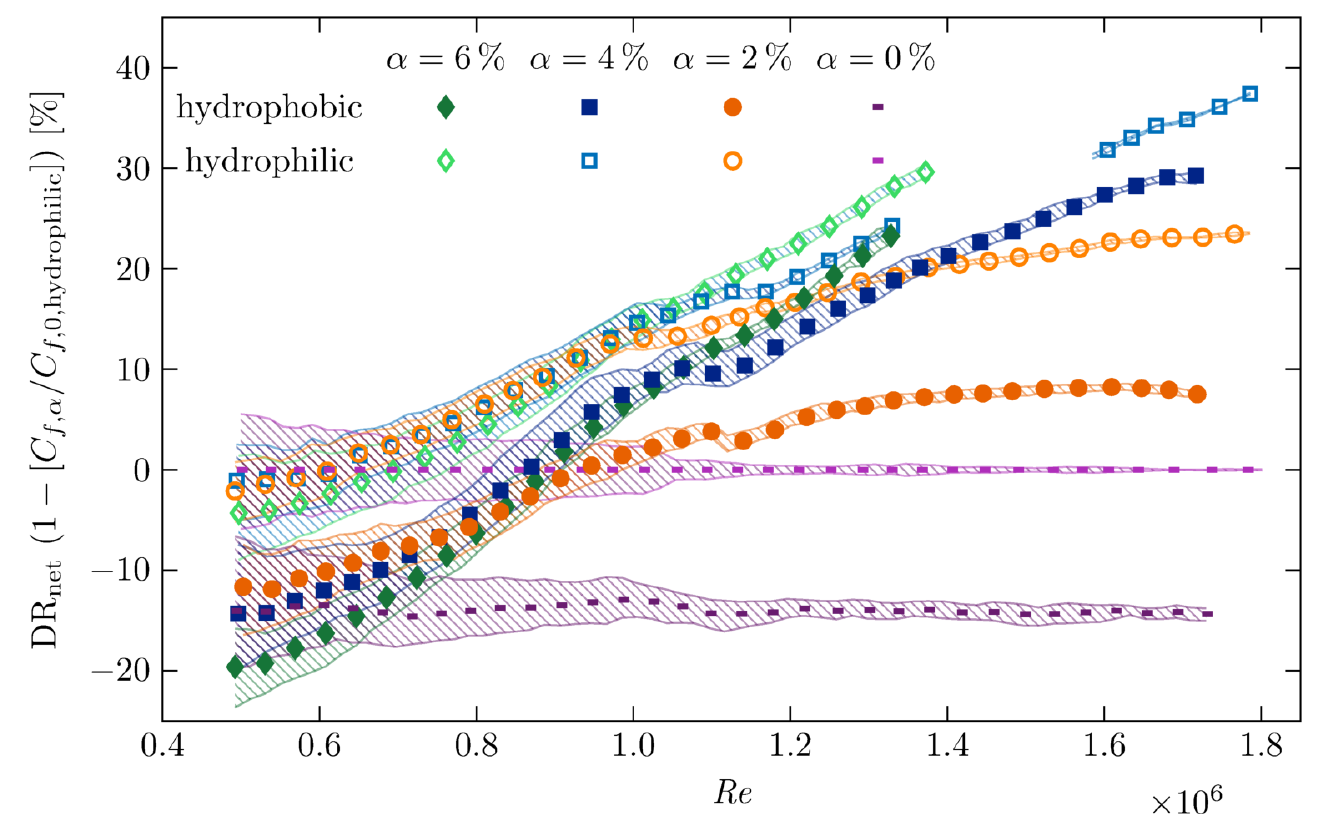}
\caption{Plot of the net drag reduction (DR$_\text{net}$) based on skin friction coefficient $C_f$ (equation~(\ref{eq:DR2})) versus $\Rey$. Compared to figure~\ref{fig:DR1}, is the drag reduction here determined using the hydrophilic IC with $\alpha = 0$ as the reference ($C_{f,0,\text{hydrophilic}}$), whereas in figure~\ref{fig:DR1} the reference is the same IC as used for the $\alpha > 0$ measurement (so either hydrophobic or hydrophilic) with $\alpha = 0$.  For all values of $\alpha$, the use of the coating results in less efficient net DR compared to an uncoated cylinder. Only every second data point is shown to improve readability of the plot. The shaded regions represent the spread in data. \add{The size of the error bars based on the accuracy of the torque sensor is smaller than the marker size.} Due to heavy vibrations in the setup resulting from a non-symmetric distribution of air, no data was acquired in the region between $\Rey = 1.3\times10^6$ and $\Rey = 1.6\times10^6$ for $\alpha=\perc{4}$ and for $\Rey > 1.4\times10^6$ when $\alpha=\perc{6}$.}\label{fig:DR2}
\end{figure}

The definition of DR$_\text{net}$ in equation~(\ref{eq:DR2}) is used in figure~\ref{fig:DR2} to study the combined effect of both air bubbles ($\alpha > \perc{0}$) and a {hydrophobic} IC, using the hydrophilic IC with $\alpha = \perc{0}$ as a reference for all cases. Compared to figure~\ref{fig:DR1}, all data obtained using the {hydrophobic} IC is shifted downwards by roughly $\perc{15}$.  %Again, a difference is seen between $\alpha \geq \perc{4}$ and $\alpha < \perc{4}$.
For the lines $\alpha = \perc{0}$ and $\alpha = \perc{2}$, the shift in DR$_\text{net}$ is nearly constant, whereas for $\alpha = \perc{4}$ and $\alpha = \perc{6}$ the difference in DR$_\text{net}$ goes down with increasing $\Rey$, owing to the same competing mechanisms as discussed previously. Nonetheless it is clear from figure~\ref{fig:DR2} that for all values of $\alpha$, the use of a {hydrophobic} IC results in a less efficient net DR compared to a hydrophilic IC. The introduction of the {hydrophobic} coating adds roughness to the otherwise hydrodynamically smooth hydrophilic surface, which in this case explains the significant change in the net drag force.

\subsection{Repeatability and measurement errors of torque measurements}
For the highest Reynolds number achieved in this research, namely $\Rey = 1.8\times10^6$, the shear stress at the surface $\tau_w = \mathcal{T} / 2 \pi r_i^2 L_\text{mid}$ is \SI{274}{\Pa}. To ensure that the {hydrophobic} coating applied to the IC is not adversely affected by this high shear stress, we measured the $\alpha = \perc{0}$ reference measurement twice after every series of $\alpha > \perc{0}$ measurements, as indicated in table~\ref{table:measurement_order}. The skin friction coefficients found in the $\alpha = \perc{0}$ measurements have a spread of \perc{3}. This spread shows no trend, which suggests that the properties of the coating remain constant throughout the experiments. This was also confirmed by visual inspection of the coating and by comparison of SEM images of the coating taken before and after the measurements.  \\
The repeated measurements of $\alpha = \perc{2}, \perc{4}, \text{and } \perc{6}$ show a spread in skin friction coefficient of $\perc{2}, \perc{4}, \text{and } \perc{2}$, respectively, again suggesting good measurement repeatability. \add{The accuracy of the torque sensor is rated by the manufacturer to be $\pm \perc{0.25}$ of its maximum rated output. Error regression analysis has shown that the largest error in figures~\ref{fig:DR1}~and~\ref{fig:DR2} is slightly smaller than the markers used.} The data spread is reflected by the shaded regions in figures~\ref{fig:DR1}~and~\ref{fig:DR2}. These regions are derived by using the minimum and maximum values from the repeated measurements for $C_{f,0}$ and $C_{f,\alpha}$.

\subsection{Velocity profiles}
\begin{figure}
\centering \includegraphics{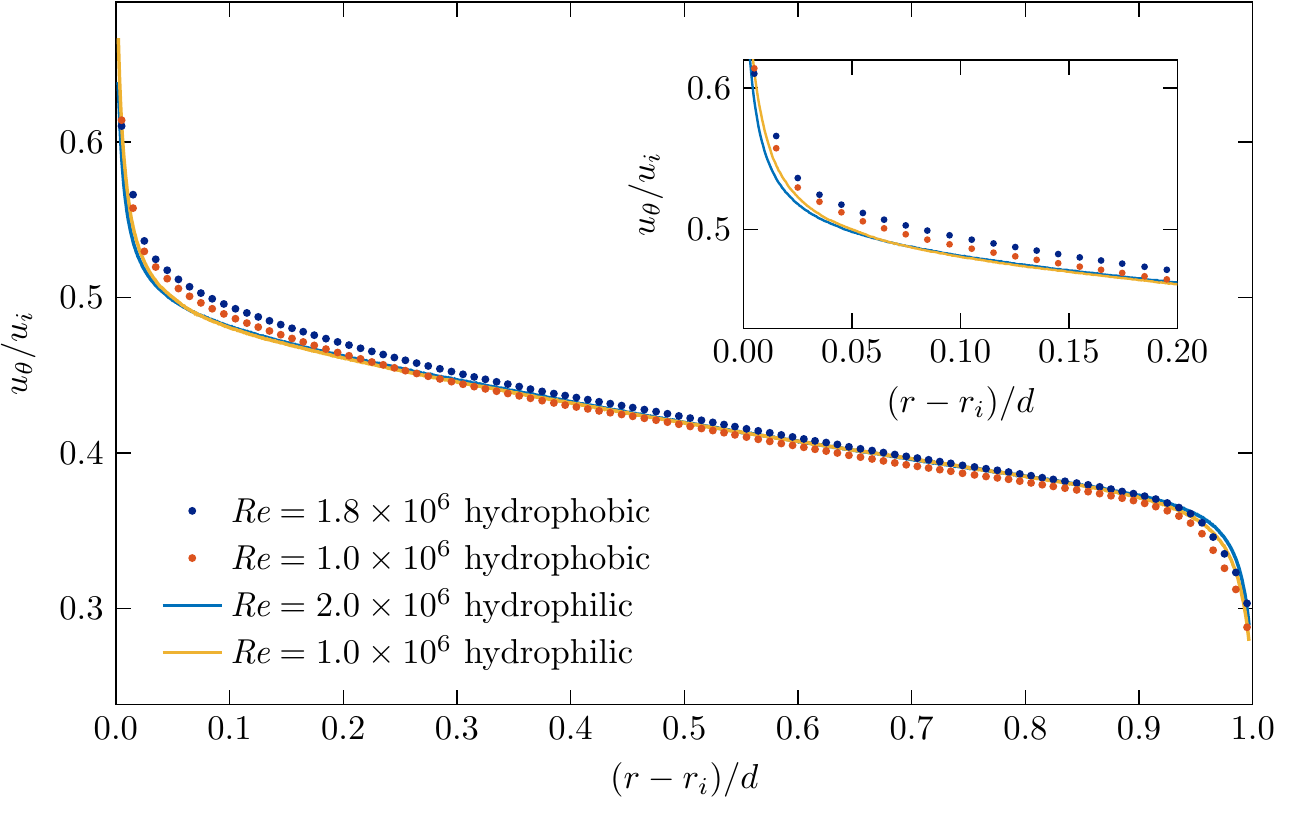} \caption{Plot of azimuthal velocity normalized with inner cylinder velocity $u_\theta / u_i$ versus the normalized gap width between inner and outer cylinder $(r-r_i) / d$. The working fluid is without air, so $\alpha = \perc{0}$. Compared is the rough \textit{hydrophobic} coating with the smooth \textit{hydrophilic} steel inner cylinder. The \textit{hydrophilic} data is provided by~\cite{hui13} using the same experimental setup. The inset shows the region close to the inner cylinder. }\label{fig:PIVplot}
\end{figure}

Shown in figure~\ref{fig:PIVplot} are the results on the velocity profiles from the PIV measurements. Plotted are the azimuthal velocities normalized with the velocity of the inner cylinder $u_\theta / u_i$, versus the normalized position between the inner and outer cylinder $(r - r_i) / d$, for different Reynolds numbers. We compare the \textit{hydrophilic} IC to the \textit{hydrophobic} IC, for a single-phase flow with $\alpha = \perc{0}$. A clear difference is seen in the region close to the IC, with larger velocities for the {hydrophobic} IC compared to the hydrophilic IC at similar Reynolds numbers. This is attributed to the larger roughness of the {hydrophobic} IC compared to the hydrophilic IC. A rougher surface can transport more energy to the flow compared to a smooth surface, resulting in larger velocities in the near wall region for the same driving of the flow~\citep{Zhu2018}. In other flow configurations, when the wall is not used to drive the flow, lower velocities will be found for more rough surfaces~\citep{Flack2010,MacDonald2016,Busse2017}. For larger $\Rey$ we find larger velocities close to the IC, indicating that the effect of the roughness is stronger with larger $\Rey$. This reflects the reduction in thickness of the viscous sublayer, which results in a larger fraction of the small scale roughness (pores) of the {hydrophobic} IC becoming a relevant length scale to the flow.

%% file: summary_conclusions.tex
% !TEX root = Main document.tex

\section{Summary and conclusions}
The influence of a {hydrophobic} wall on drag reduction was studied in a highly turbulent Taylor--Couette flow. We applied a {hydrophobic} coating to the otherwise smooth and hydrophilic inner cylinder (IC) of the Taylor--Couette setup. In single-phase flow, we found a constant increase in drag of about \perc{14} for the rough hydrophobic wall compared to the smooth hydrophilic wall over the whole range of Reynolds numbers $5.0\e{5} \leq \Rey \leq 1.8\e{6}$ measured. For bubbly two-phase flow however, the addition of air bubbles to the flow resulted in more drag reduction for the rough hydrophobic IC as compared to the hydrophilic IC, using the same volume fraction of air bubbles $\alpha$. For $\alpha \geq 4\%$, more DR was found over nearly the whole range of $\Rey$. Only in the region of $\Rey < 7.0\e{5}$ -- where the measurement uncertainty is highest -- a slight drag increase was found for the {hydrophobic} IC when $\alpha = 6\%$. A strong difference in DR behaviour is found when comparing $\alpha = 2\%$ with $\alpha \geq 4\%$. The void fraction $\alpha = 2\%$ gives a clear drag increase above $\Rey = 10^6$, indicating that the bubble drag mechanism is more effective with a superhydrophic IC when sufficient air bubbles are present in the flow.
This can be explained by the micro-scale surface geometry of the surface, acting as roughness to the flow and hence increasing the drag. The effect of the {hydrophobic} coating is therefore twofold: (i) a more effective bubble drag reduction mechanism, and (ii) an increase in drag from the surface roughness. The role of roughness is confirmed by comparing drag measurements of the {hydrophobic} IC to drag measurements of the smooth hydrophilic IC with $\alpha = 0$, which shows more drag for all values of $\alpha$.

The effect of roughness is more pronounced with larger $\Rey$, since the thickness of the viscous sublayer is then smaller, making it compatible with the roughness length scales. This is confirmed by velocity profile measurements, showing that the normalized azimuthal flow velocities near the {hydrophobic} IC are larger compared to the smooth hydrophilic IC for the same $\Rey$. This indicates that for larger $\Rey$ the {hydrophobic} IC appears rougher for the flow.

Whereas the drag continues to increase with $\Rey$ for the {hydrophobic} IC compared to the hydrophilic IC when $\alpha < 4\%$, the difference in drag appears to level off for $\alpha \geq 4\%$, showing a much weaker dependence on $\Rey$. Apparently, above a certain $\Rey$ with $\alpha \geq 4\%$, the difference between the two competing effects reaches a constant value. The result is a constant increase in DR for the {hydrophobic} IC. The difference in DR does also not vary significantly with $\alpha$ for $\alpha \geq 4\%$. This leads us to the conclusion that, although a minimum amount of air is required for the {hydrophobic} coating to provide more effective bubble DR, adding more air beyond this minimum barrier will not necessarily result in more drag reduction. We hope that our work will give guidelines for industrial applications of bubbly drag reduction in hydrophobic wall-bounded turbulence, such as in naval applications or in pipelines.